\documentclass[aps,pra,twocolumn,superscriptaddress]{revtex4-1}

\usepackage{graphicx}
\usepackage{amsmath}
\usepackage{marvosym}
\newcommand*\myat{{\fontfamily{ptm}\selectfont @}}

\begin{document}


\title{Evolution of Photon Beams through a Nested Mach-Zehnder Interferometer using Classical States of Light}



\author{Anarta Roy}
\email{anarta.iiserb\myat gmail.com}
\affiliation{Department of Physics, Indian Institute of Science Education and Research, Bhopal Bypass Road, Bhopal 462 066, India}

\author{Sibasish Ghosh}
\email{sibasish\myat imsc.res.in}
\affiliation{Optics \& Quantum Information Group, The Institute of Mathematical Sciences, HBNI, C.I.T. Campus, Taramani, Chennai 600 113, India}



\date{\today}

\begin{abstract}
In this paper, we present a coherent state-vector method which can explain the results of a nested linear Mach-Zehnder Interferometric experiment. Such interferometers are used widely in Quantum Information and Quantum Optics experiments and also in designing quantum circuits. We have specifically considered the case of an experiment by       Danan \emph{et al.} (Phys. Rev. Lett. \textbf{111}, 240402 (2013)) where the outcome of the experiment was spooky by our intuitive guesses. However we have been able to show by our method that the results of this experiment is indeed expected within the standard formalism of Quantum Mechanics using any classical state of a single-mode radiation field as the input into the nested interferometric set-up of the aforesaid experiment and thereby looking into the power spectrum of the output beam. 
\end{abstract}


\maketitle


\section{INTRODUCTION}

Quantum Mechanics can explain various counter-intuitive phenomena of the subatomic world \cite{dirac1981principles}. An interferometer is generally used to describe several counter-intuitive features one gets to see in Quantum Mechanics \cite{dirac1981principles}. 

In a fairly recent experiment by Danan \emph{et al.}, the results were spooky (ref. \cite{danan2013asking}). The experiment involves a nested Mach-Zehnder Interferometer(MZI) inside which, mirrors A, B, C, E, and F vibrate at different frequencies (see FIG. 1). This interferometric arrangement allows the passage of light (represented as lines with arrows in FIG. 1) through several beam splitters namely BS1, BS2, BS3, and BS4 and via the vibrating mirrors A, B, C, E, and F. The photons of light record the frequencies of the vibrating mirrors (via which they pass). The final output beam of light from beam splitter BS4 remains directed towards a quad-cell detector D, which records the frequencies (in the form of a power spectrum) of the vibrating mirrors via which the photons of light have passed. This experiment \cite{danan2013asking} seems to infer about the path the photon has taken while passing through the interferometer. 

The results were explained by the authors using the two-state vector formalism (TSVF) and weak values in QM \cite{aharonov2002two,aharonov1964time,PhysRevA.41.11}. This explanation was followed by a series of arguments and counter-arguments regarding how the outcome of this experiment should be analysed \cite{salih2014comment,fu2015ideal,huang2014comments,svensson2014comments,wiesniak2014criticism,griffiths2016particle}. All these arguments have dealt with the methods which might be employed in a search for the path of photons through a nested Mach-Zehnder Interferometer (MZI) \cite{danan2013asking}. Two theoretical approaches have been employed for explaining the results of this experiment. One uses a classical optics formalism \cite{saldanha2014interpreting} and the other employs a ``one-state vector quantum-mechanical" approach \cite{bartkiewicz2015one}.

\begin{figure}[h!]
\centering
\includegraphics[width=9cm,height=6.5cm]{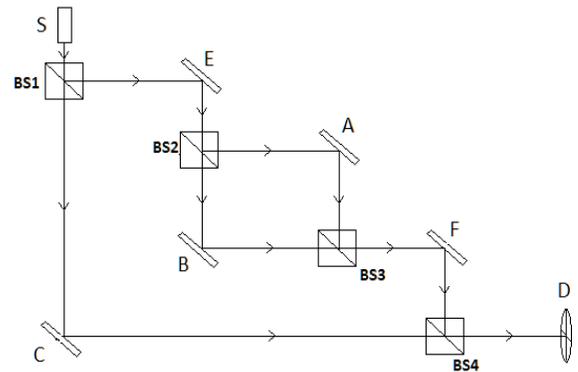}
\caption{Mach-Zehnder Interferometer(nested) consisting of beam splitters BS1, BS2, BS3, BS4 and mirrors A, B, C, E and F. The arrows show the path of light beam from source S to detector D \cite{danan2013asking}.} 
\end{figure} 

In this paper, we will analyse the outcome of this experiment using a standard coherent-state formulation. Our results and methods will be able to explain the outcome of this experiment and will be able to shed some more light on this topic \cite{danan2013asking}. Given that a coherent state is mainly a classical state, we can say with confidence that this experiment falls in the classical regime. However, we have followed a purely quantum mechanical way to look at this problem.

In section II we will discuss how a quantum mechanical lossless beam-splitter transforms its input modes to output modes \cite{campos1989quantum}. In sections III and IV we will discuss the evolution of the system-states as they pass through different beam-splitters and mirrors in this experiment \cite{danan2013asking}. In section III, we analyze setup 1 and in section IV we analyze setup 2 of this experiment \cite{danan2013asking}. In section V we will discuss the power spectrum of the output states for setups 1 and 2. FIG. 1 gives a schematic representation of setup 1 of this experiment. Finally we conclude in section VI.

\section{General lossless beam splitter transformation}

\begin{figure}[h!]
\includegraphics[width=8cm,height=6cm]{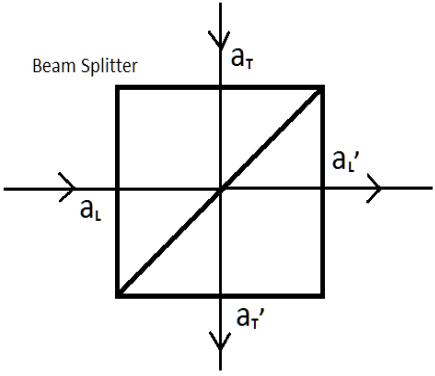}
\caption{A generic quantum mechanical lossless beam splitter with input modes $\hat{a}_{L}$ \& $\hat{a}_{T}$ and output modes $\hat{a}_{L}^{\prime}$ \& $\hat{a}_{T}^{\prime}$.}  
\end{figure}

There are two input modes and correspondingly two output modes in a quantum-mechanical lossless beam splitter. For the beam-splitter representation in Fig. 2, the input modes are represented by annihilation operators $\hat{a}_{L}$ \& $\hat{a}_{T}$. The output modes are given by corresponding annihilation operators $\hat{a}_{L}^{\prime}$ \& $\hat{a}_{T}^{\prime}$. They are related by the transformation matrix $\hat{B}$ through the following equation. Please see Ref. \cite{campos1989quantum}:

\begin{equation} 
\begin{pmatrix}
\hat{a}_{T}^{\prime}  \\
\hat{a}_{L}^{\prime}
\end{pmatrix}
=
\hat{B}
\begin{pmatrix}
\hat{a}_{T} \\
\hat{a}_{L}
\end{pmatrix}
=
\begin{pmatrix}
B_{11}  &  B_{12} \\
B_{21}  &  B_{22} 
\end{pmatrix}
\begin{pmatrix}
\hat{a}_{T} \\
\hat{a}_{L}
\end{pmatrix}
\label{eq:eqn1}
\end{equation}

The unitary operator $U_{B}$ of the transformation characterizes the beam-splitter. Following the angular momentum theory, $U_{B}$ is the two-dimensional representation of the rotation group (SO(3)) \cite{campos1989quantum}.  

\begin{equation}
U_{B}(\Phi,\Theta,\Psi) = e^{-i\Phi \hat{L}_{3}} e^{-i\Theta \hat{L}_{2}} \\ e^{-i\Psi \hat{L}_{3}}
\end{equation}

From (2), we can derive the expression for matrix $\hat{B}$ \cite{campos1989quantum}, which is

\begin{equation}
\hat{B}
=
\begin{pmatrix}
\cos(\Theta /2)e^{i([\Psi + \Phi]/2)} & \sin(\Theta /2)e^{i([\Psi -  \Phi]/2)} \\
-\sin(\Theta /2)e^{-i([\Psi - \Phi]/2)} & \cos(\Theta /2)e^{-i([\Psi + \Phi]/2)}
\end{pmatrix}
\label{eq:eqn3}
\end{equation}
where $\Phi, \Theta, \Psi$ are quantum mechanical analogues of the classical Euler angles. 

Now we consider a general input state as $\left| \psi_{in}\right\rangle$ which is a tensor product of two fock states. We use subscripts T and L to denote states entering the Top port and the Left port of the beam splitter respectively.

\begin{equation}
\left| \psi_{in}\right\rangle = \left| n \right\rangle_{T} \left| m \right\rangle_{L} = (n!m!)^{-1/2}[(\hat{a}_{T}^{\dagger})^{n} \otimes (\hat{a}_{L}^{\dagger})^{m}]\left| 0 \right\rangle \left| 0 \right\rangle
\end{equation}

Also from (1), we get (noting that $\hat{B}$ is invertible and $B^{\dagger}_{ij} = B^{*}_{ij}$),

\begin{equation}
\begin{pmatrix}
\hat{a}_{T} \\
\hat{a}_{L}
\end{pmatrix}
= 
\begin{pmatrix}
B_{11}^{\dagger}  &  B_{21}^{\dagger} \\
B_{12}^{\dagger}  &  B_{22}^{\dagger} 
\end{pmatrix}
\begin{pmatrix}
\hat{a}_{T}^{\prime}  \\
\hat{a}_{L}^{\prime}
\end{pmatrix}
\label{eq:eqn5}
\end{equation}

Hence, in terms of output annihilation operators (modes)                    $\hat{a}_{L}^{\prime}$ \& $\hat{a}_{T}^{\prime}$, the input modes can be expressed as,

\begin{eqnarray}
\hat{a}_{T} = B_{11}^{\dagger}\hat{a}_{T}^{\prime} + B_{21}^{\dagger}\hat{a}_{L}^{\prime} \\
\hat{a}_{L} = B_{12}^{\dagger}\hat{a}_{T}^{\prime} + B_{22}^{\dagger}\hat{a}_{L}^{\prime}
\label{eq:eqn6,7}
\end{eqnarray}

Putting (6) and (7) in expression (4), 

\begin{eqnarray}
\begin{aligned}
\left| \psi_{in}\right\rangle = (n!m!)^{-1/2}[(\hat{a}_{T}^{\dagger})^{n} \otimes (\hat{a}_{L}^{\dagger})^{m}]\left| 0 \right\rangle \left| 0 \right\rangle  \\
= (n!m!)^{-1/2}[((B_{11}^{\dagger}\hat{a}_{T}^{\prime} +  B_{21}^{\dagger}\hat{a}_{L}^{\prime})^{\dagger})^{n} \otimes \\ ((B_{12}^{\dagger}\hat{a}_{T}^{\prime} +  B_{22}^{\dagger}\hat{a}_{L}^{\prime})^{\dagger})^{m}] 
\left| 0 \right\rangle \left| 0 \right\rangle  
\end{aligned}
\end{eqnarray}

After some rearrangement with expression (8), we get $\left| \psi_{out}\right\rangle$.

\begin{eqnarray}
\begin{aligned}
\left| \psi_{out}\right\rangle = (n!m!)^{-1/2}\sum\limits_{j=0}^{n} \sum\limits_{k=0}^{m}
\begin{pmatrix}
n \\
j
\end{pmatrix}
\begin{pmatrix}
m \\
k
\end{pmatrix} \\
[(n-j+m-k)!(j+k)!]^{1/2} B_{11}^{n-j}B_{21}^{j}\\B_{12}^{m-k}B_{22}^{k} \left| n-j+m-k \right\rangle_{T^{\prime}}\left| j+k \right\rangle_{L^{\prime}}
\end{aligned}
\end{eqnarray}

We note that this output state has been obtained after applying the unitary transformation corresponding to $\hat{B}$ on the input state. This gives the most general transformation of the input state to the output state of a quantum mechanical lossless beam splitter. From (3), we have the values of the matrix elements of $\hat{B}$.

\begin{eqnarray}
\begin{aligned}
B_{11} = \cos(\Theta /2)e^{i([\Psi + \Phi]/2)} \\
B_{12} = \sin(\Theta /2)e^{i([\Psi -  \Phi]/2)} \\
B_{21} = -\sin(\Theta /2)e^{-i([\Psi - \Phi]/2)} \\
B_{22} = \cos(\Theta /2)e^{-i([\Psi + \Phi]/2)}
\end{aligned}
\end{eqnarray}

From the above elements we get the transmittance $\tau$ and reflectance $\rho = (1-\tau)$ of the beam splitter \cite{campos1989quantum}.

\begin{eqnarray}
\begin{aligned}
\left|B_{11}\right|^{2} = \left|B_{22}\right|^{2} = \tau = \cos^{2}(\Theta/2)  \\
\left|B_{12}\right|^{2} = \left|B_{21}\right|^{2} = \rho = \sin^{2}(\Theta/2)  \\
\end{aligned}
\end{eqnarray}
Now we are at a position to begin the analysis of the experimental results of Danan \emph{et al}.

\section{ANALYSIS OF SETUP 1}

Here, we discuss the first setup of the experiment (setup 1). This is given in the diagrammatic representation in Fig 3. We will be analyzing each and every beam-splitter and mirror, one after the other as the beam passes through each of them. 

\begin{figure}[h!]
\includegraphics[width=9.3cm,height=7.5cm]{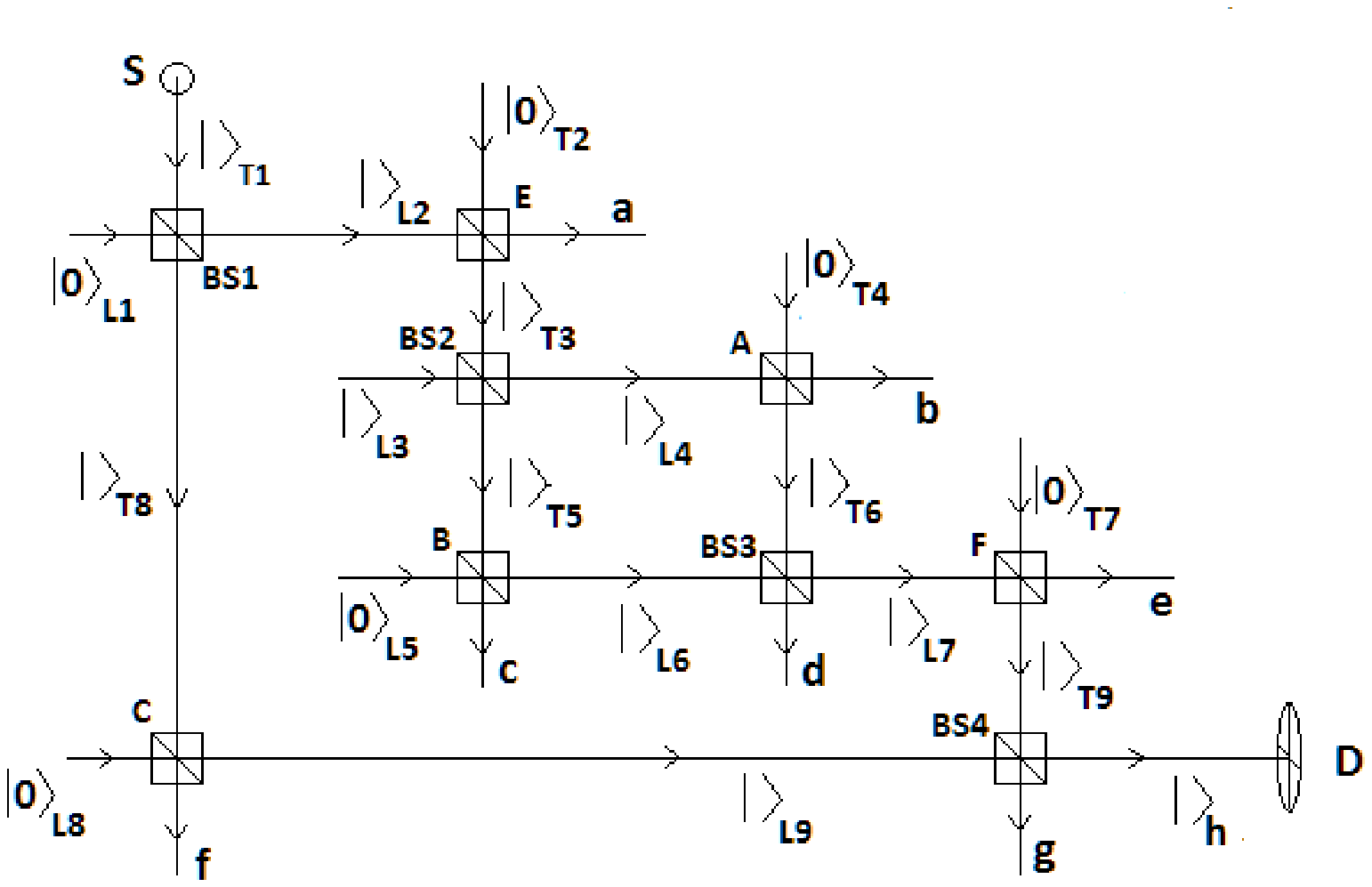}
\caption{Setup 1 of \cite{danan2013asking} showing state vectors of all the photon beams. We have to follow the state vectors to understand the evolution of the states from S to D. The power spectrum from detector D records the frequencies of oscillations of all the five mirrors A,B,C,E and F. The final state reaching detector D is $\left|\right\rangle_{h}$.}  
\end{figure}

\subsection{Beam Splitter 1}

The action of the first beam splitter (BS1) in FIG. 3 is governed by the general scheme mentioned before in Section II. The same scheme also governs the other beam splitters in the figure. Let $\left|\alpha\right\rangle$ be a coherent state of a single mode system. We consider the input state as,

\[
\begin{aligned}
\left| \psi_{in}\right\rangle_{BS1} = \left|\alpha\right\rangle_{T1} \otimes \left|0\right\rangle_{L1} \\ = e^{-|\alpha|^{2}/2}\sum\limits_{n=0}^{\infty} \frac{\alpha^{n}}{\sqrt{n!}} 
\left|n\right\rangle_{T1} \otimes \left|0\right\rangle_{L1} \\
= e^{-|\alpha|^{2}/2}\sum\limits_{n=0}^{\infty} \frac{\alpha^{n}}{\sqrt{n!}}
[(n!)^{-1/2}\sum\limits_{j=0}^{n}
\begin{pmatrix}
n \\
j
\end{pmatrix} \\ [(n-j)!j!]^{1/2} B_{11}^{n-j} B_{21}^{j} \left|n-j\right\rangle_{T8} \left|j\right\rangle_{L2}
\end{aligned}
\]
We have given labels T8 and L2 to the output states of BS1 (refer to Fig 3). According to eqn (3) of Section II, $\Phi_{1}$ and $\Psi_{1}$ are the constant Euler angles for BS1. Working on the above expression, the final state is obtained as 

\begin{equation}
\begin{aligned}
\left| \psi_{out}\right\rangle_{BS1} = \left|\alpha\tau e^{i(\frac{\Psi_{1}+\Phi_{1}}{2})}\right\rangle_{T8} \otimes \\ \left|-\alpha\rho e^{-i(\frac{\Psi_{1}-\Phi_{1}}{2})}\right\rangle_{L2}
\end{aligned}
\end{equation}
a tensor product of two coherent states. Noting that $\tau = \sqrt{\frac{1}{3}}$ and $\rho = \sqrt{\frac{2}{3}}$ for BS1 (1:2 beam-splitting as used in ref. \cite{danan2013asking}), the output state is given by
\begin{equation}
\begin{aligned}
\left| \psi_{out}\right\rangle_{BS1} = \left|\alpha \sqrt{\frac{1}{3}} e^{i(\frac{\Psi_{1}+\Phi_{1}}{2})}\right\rangle_{T8} \otimes \\ \left|-\alpha \sqrt{\frac{2}{3}} e^{-i(\frac{\Psi_{1}-\Phi_{1}}{2})}\right\rangle_{L2}
\end{aligned}
\end{equation}
System T8 now goes towards mirror C and L2 towards mirror E.

\subsection{Mirror E}

The action of a mirror is the same as that of a beam splitter which has $\tau = 0$ and $\rho = 1$. The input state for mirror E is given by
\begin{equation}
\left| \psi_{in}\right\rangle_{E} = \left|0\right\rangle_{T2} \otimes \left|-\alpha \sqrt{\frac{2}{3}} e^{-i(\frac{\Psi_{1}-\Phi_{1}}{2})}\right\rangle_{L2}
\end{equation}
Working on this state, we obtain the output to be
\begin{equation}
\begin{aligned}
\left| \psi_{out}\right\rangle_{E} = \left|\alpha_{E}B_{12}\right\rangle_{T3} \otimes \left|\alpha_{E}B_{22}\right\rangle_{a}
\end{aligned}
\end{equation}
where $\alpha_{E} = -\alpha \sqrt{\frac{2}{3}} e^{-i(\frac{\Psi_{1}-\Phi_{1}}{2})}$.

All the mirrors in this experiment vibrate with different frequencies and same amplitudes. Hence the phases of the beam are time-dependent. We have thus (see equations (2) and (3)),
\begin{center}
$B_{12} = e^{i[(\Psi_{E}(t)-\Phi_{E}(t))/2]}$,
$B_{22} = 0$
\end{center}
Here $\Psi_{E}(t)$ and $\Phi_{E}(t)$ are the time-dependent Euler angular phase factors for mirror E given by:
\begin{equation}
\begin{aligned}
\Psi_{E}(t) = \psi_{0}\sin (2\pi f_{E} t) \\
\Phi_{E}(t) = \phi_{0}\sin (2\pi f_{E} t) \\
\end{aligned}
\end{equation}
where $f_{E}$ (in Hz units) denotes the frequency of oscillation for mirror E. $\psi_{0}$ and $\phi_{0}$ are constant amplitudes which are same for all the mirrors in ref. \cite{danan2013asking}.
Thus, the final state from mirror E is
\begin{equation}
\begin{aligned}
\left| \psi_{out}\right\rangle_{E} = \left|\alpha_{E}e^{i[(\Psi_{E}(t)-\Phi_{E}(t))/2]}\right\rangle_{T3} \otimes \left|0\right\rangle_{a} 
\end{aligned}
\end{equation}
System T3 is now directed towards BS2.

\subsection{Beam Splitter 2}

BS2 is a 1:1 beam splitter as described in the experiment \cite{danan2013asking}. This means $\tau = 1/2$ \& $\rho = 1/2$. The input state to BS2 is given by:
\begin{equation}
\begin{aligned}
\left| \psi_{in}\right\rangle_{BS2} = \left|\alpha_{E}e^{i[(\Psi_{E}(t)-\Phi_{E}(t))/2]}\right\rangle_{T3} \otimes \left|0\right\rangle_{L3} 
\end{aligned}
\end{equation}
If we take $\alpha^{\prime}_{E} = \alpha_{E}e^{i[(\Psi_{E}(t)-\Phi_{E}(t))/2]}$, the output state from BS2 is found to be
\begin{equation}
\begin{aligned}
\left| \psi_{out}\right\rangle_{BS2} = \left|\alpha^{\prime}_{E}B_{11}\right\rangle_{T5}\left|\alpha^{\prime}_{E}B_{21}\right\rangle_{L4} \\
= \left|\alpha_{E}e^{i[(\Psi_{E}(t)-\Phi_{E}(t))/2]}\sqrt{\frac{1}{2}}e^{i(\Psi_{2}+\Phi_{2})/2}\right\rangle_{T5} \otimes \\ \left|-\alpha_{E}e^{i[(\Psi_{E}(t)-\Phi_{E}(t))/2]}\sqrt{\frac{1}{2}}e^{-i(\Psi_{2}-\Phi_{2})/2}\right\rangle_{L4}
\end{aligned}
\end{equation}
As in the previous beam splitter, here $\Phi_{2}$ and $\Psi_{2}$ are the constant Euler angles for the beam splitter BS2. $\alpha_{E}$ is known from mirror E. The system T5 goes towards the mirror B, while system L4 goes towards mirror A. We note here, that the vibration frequency of mirror E is stored in both the states of T5 and L4 (see equation (18)). 

\subsection{Mirror A}

System L4 enters Mirror A. We thus have here, according to equation (18), 
\[
\alpha_{A} = -\alpha_{E}e^{i[(\Psi_{E}(t)-\Phi_{E}(t))/2]}\sqrt{\frac{1}{2}}e^{-i(\Psi_{2}-\Phi_{2})/2}
\]
Thus the input state to the mirror A is given by:
\begin{equation}
\begin{aligned}
\left| \psi_{in}\right\rangle_{A} = \left|0\right\rangle_{T4}\ \otimes \left|\alpha_{A}\right\rangle_{L4} 
\end{aligned}
\end{equation}
After the action of the mirror, state is given by:
\begin{equation}
\begin{aligned}
\left| \psi_{out}\right\rangle_{A} = \left|\alpha_{A}B_{12}\right\rangle_{T6} \otimes \left|\alpha_{A}B_{22}\right\rangle_{b}
\end{aligned}
\end{equation}
We take the time dependent phases for mirror A as $\Phi_{A}(t)$ \& $\Psi_{A}(t)$,
\begin{equation}
\begin{aligned}
\left| \psi_{out}\right\rangle_{A} = \left|\alpha_{A}e^{i[(\Psi_{A}(t)-\Phi_{A}(t))/2]}\right\rangle_{T6} \otimes \left|0\right\rangle_{b} 
\end{aligned}
\end{equation}
where ($f_{A}$ in Hz units is the oscillation frequency for mirror A; $\psi_{0}$ and $\phi_{0}$ are its constant amplitudes of oscillation)
\begin{equation}
\begin{aligned}
\Psi_{A}(t) = \psi_{0}\sin (2\pi f_{A} t) \\
\Phi_{A}(t) = \phi_{0}\sin (2\pi f_{A} t) \\
\end{aligned}
\end{equation}
System T6 is directed towards beam-splitter 3 or BS3. Before moving on to BS3, we will first look at mirror B.

\subsection{Mirror B}

Similar to the case of mirror A, the system T5 now enters the Mirror B. We thus take here 
\[
\alpha_{B} = \alpha_{E}e^{i[(\Psi_{E}(t)-\Phi_{E}(t))/2]}\sqrt{\frac{1}{2}}e^{i(\Psi_{2}+\Phi_{2})/2}
\]
This is according to equation (19), for system T5, where $\alpha_{B}$ is coherent. The input state to the mirror B is
\begin{equation}
\begin{aligned}
\left| \psi_{in}\right\rangle_{B} = \left|\alpha_{B}\right\rangle_{T5} \otimes \left|0\right\rangle_{L5} 
\end{aligned}
\end{equation}
After the action of the mirror B on $\left| \psi_{in}\right\rangle_{B}$, the output state is given by
\begin{equation}
\begin{aligned}
\left| \psi_{out}\right\rangle_{B} = \left|\alpha_{B}B_{11}\right\rangle_{c} \otimes \left|\alpha_{B}B_{21}\right\rangle_{L6}
\end{aligned}
\end{equation}
See equations (2) and (3). We take the time dependent phases for mirror B as $\Psi_{B}(t)$ and $\Phi_{B}(t)$ where ($f_{B}$ in Hz units),
\begin{equation}
\begin{aligned}
\Psi_{B}(t) = \psi_{0}\sin (2\pi f_{B} t) \\
\Phi_{B}(t) = \phi_{0}\sin (2\pi f_{B} t) \\
\end{aligned}
\end{equation}
This is similar to the previous mirror A. The constant amplitudes are the same as in mirror A. The oscillation amplitudes for all the mirrors in ref. \cite{danan2013asking} are constant and equal. Hence all amplitudes of oscillations for the mirrors in ref. \cite{danan2013asking} are given by $\psi_{0}$ and $\phi_{0}$. $f_{B}$ is the oscillation frequency for mirror B.
\begin{equation}
\begin{aligned}
\left| \psi_{out}\right\rangle_{B} = \left|0\right\rangle_{c} \otimes \left|-\alpha_{B}e^{-i[(\Psi_{B}(t)-\Phi_{B}(t))/2]}\right\rangle_{L6} 
\end{aligned}
\end{equation}
The system L6 is also directed towards the beam splitter BS3. Now, we will analyse the case of BS3. 

\subsection{Beam Splitter 3}

Before discussing the BS3 (50:50) transformation, we look at the unitary transformation corresponding to any beam splitter action on an input state of the form $\left|0\right\rangle_{T} \otimes \left|\gamma\right\rangle_{L}$, where $\left|\gamma\right\rangle$ is a single mode coherent state:

\begin{equation}
U_{BS}\left|0\right\rangle_{T} \otimes \left|\gamma\right\rangle_{L} = \left|\gamma B_{12}\right\rangle_{T^{\prime}} \otimes \left|\gamma B_{22}\right\rangle_{L^{\prime}}
\end{equation}
This unitary transformation follows from equation (9) for a coherent state $\left|\gamma\right\rangle$. Hence, if we know the RHS in (25), we are in a position to find out $\left|\gamma\right\rangle$. We apply this method for BS3. Here,
\begin{equation}
\begin{aligned}
\alpha^{\prime}_{A} = \alpha_{A}e^{i[(\Psi_{A}(t)-\Phi_{A}(t))/2]}, \\
\alpha^{\prime}_{B} = -\alpha_{B}e^{-i[(\Psi_{B}(t)-\Phi_{B}(t))/2]},
\end{aligned}
\end{equation}
and we get,
\begin{equation}
\begin{aligned}
\left| \psi_{in}\right\rangle_{BS3} = \left|\alpha^{\prime}_{A}\right\rangle_{T6} \otimes \left|\alpha^{\prime}_{B}\right\rangle_{L6} 
\end{aligned}
\end{equation}
The output state is found out to be
\begin{equation}
\begin{aligned}
\left| \psi_{out}\right\rangle_{BS3} = \left|0\right\rangle_{d} \otimes \left| \gamma \right\rangle_{L7}, 
\end{aligned}
\end{equation}
with
\begin{equation}
\begin{aligned}
\gamma = \alpha\sqrt{\frac{2}{3}}e^{-i[(\Psi_{1}-\Phi_{1})/2]} \times e^{i[(\Phi_{2}+\Phi_{3})/2]} \times \\ e^{i[(\Psi_{E}(t)-\Phi_{E}(t))/2]} \times e^{i[(\Psi_{A}(t)-\Phi_{A}(t))/4]} \times \\ e^{-i[(\Psi_{B}(t)-\Phi_{B}(t))/4]}.
\end{aligned}
\end{equation}
$\Phi_{3}$ is the constant Euler angle for the beam splitter BS3. Now the system L7 enters the mirror F. 

\subsection{Mirror F}

For this mirror (FIG. 3),
\begin{equation}
\left| \psi_{in}\right\rangle_{F} = \left|0\right\rangle_{T7} \otimes \left|\gamma\right\rangle_{L7},
\end{equation} 
where $\gamma$ is given in equation (32). After the mirror action, the output state is
\begin{equation}
\left| \psi_{out}\right\rangle_{F} = \left|\gamma e^{i[(\Psi_{F}(t)-\Phi_{F}(t))/2]}\right\rangle_{T9} \otimes \left|0\right\rangle_{e}
\end{equation}
where ($f_{F}$ in Hz units is the oscillation frequency for mirror F)
\begin{equation}
\begin{aligned}
\Psi_{F}(t) = \psi_{0}\sin (2\pi f_{F} t) \\
\Phi_{F}(t) = \phi_{0}\sin (2\pi f_{F} t) \\
\end{aligned}
\end{equation}
Now the system T9 goes into the beam splitter BS4.

\subsection{Mirror C}

We recall the two states that emerged from BS1. One of those states was the state of the system T8 which was directed towards mirror C. This makes the input state for mirror C as
\begin{equation}
\left| \psi_{in}\right\rangle_{C} = \left|\alpha \sqrt{\frac{1}{3}} e^{i(\frac{\Psi_{1}+\Phi_{1}}{2})}\right\rangle_{T8} \otimes \left|0\right\rangle_{L8}
\end{equation} 
By following our usual mirror action, the output state, in this case, turns out as
\begin{equation}
\left| \psi_{out}\right\rangle_{C} = \left|0\right\rangle_{f} \otimes \\ \left|-\alpha \sqrt{\frac{1}{3}}e^{i(\frac{\Psi_{1}+\Phi_{1}}{2})} e^{-i[(\Psi_{C}(t)-\Phi_{C}(t))/2]}\right\rangle_{L9}
\end{equation}
where ($f_{C}$ in Hz units is the oscillation frequency for mirror C)
\begin{equation}
\begin{aligned}
\Psi_{C}(t) = \psi_{0}\sin (2\pi f_{C} t) \\
\Phi_{C}(t) = \phi_{0}\sin (2\pi f_{C} t) \\
\end{aligned}
\end{equation}
The system L9 is now directed towards BS4, the final beam splitter in this setup.

\subsection{Beam splitter 4}

It is the last beam splitter in this setup which receives states of the systems T9 and L9, and produces the final output beam. The analysis for this transformation has been done in the same way as in BS3. The input state for this is
\begin{equation}
\begin{aligned}
\left| \psi_{in}\right\rangle_{BS4} = \left|\gamma e^{i[(\Psi_{F}(t)-\Phi_{F}(t))/2]}\right\rangle_{T9} \otimes \\ \left|-\alpha \sqrt{\frac{1}{3}}e^{i(\frac{\Psi_{1}+\Phi_{1}}{2})} e^{-i[(\Psi_{C}(t)-\Phi_{C}(t))/2]}\right\rangle_{L9}.
\end{aligned}
\end{equation}
The final state coming out of BS4 is then
\begin{equation}
\begin{aligned}
\left| \psi_{out}\right\rangle_{final} = \left| \psi_{out}\right\rangle_{BS4} = 
\left|0\right\rangle_{g} \otimes \left| \beta \right\rangle_{h}  
\end{aligned}
\end{equation}
where
\[
\begin{aligned}
\beta = \alpha e^{i(\frac{\Phi_{1}+\Phi_{4}}{2} + \frac{\Phi_{2}+\Phi_{3}}{4} + \frac{\pi}{2})}  e^{i(\frac{\Psi_{E}(t)-\Phi_{E}(t)}{4})} 
e^{i(\frac{\Psi_{F}(t)-\Phi_{F}(t)}{4})} \times \\ e^{i(\frac{\Psi_{A}(t)-\Phi_{A}(t)}{8})} e^{-i(\frac{\Psi_{B}(t)-\Phi_{B}(t)}{8})}
e^{-i(\frac{\Psi_{C}(t)-\Phi_{C}(t)}{4})}
\end{aligned}
\]
$\Phi_{4}$ is the Euler angle for the beam splitter BS4. The state of the system \emph{h} is the final state which reaches the detector D (FIG. 3). We see that the state of the system $h$ is a coherent state and in this coherent state $\left| \beta \right\rangle_{h}$, the oscillation frequencies of all the mirrors A, B, C, E, \& F are present. Also the phases for mirrors B and C have signs opposite to that of mirrors A, E and F.

Next, we will analyze the spooky results of this experiment \cite{danan2013asking}.

\section{ANALYSIS OF SETUP 2}

In setup 2 (Fig 4.) of the experiment, everything remains the same as setup 1 until we reach BS3 and then BS4. So we will analyze the outputs of only these two beam-splitters. Mirror F is not important as the beam does not reach it anymore. So we ignore it.

\begin{figure}[h!]
\includegraphics[width=9.4cm,height=7cm]{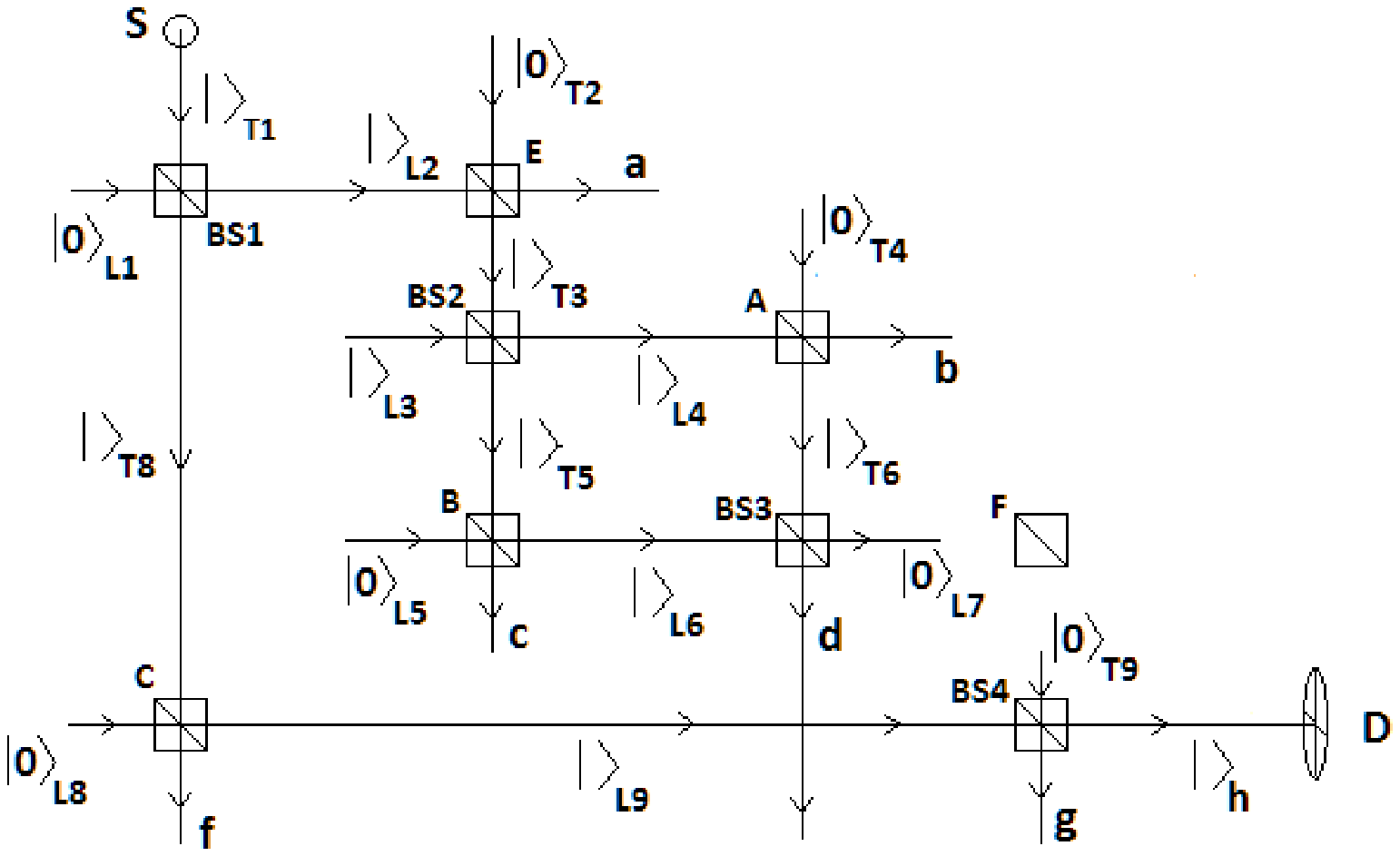}
\caption{Setup 2 \cite{danan2013asking} shows the states of the beam when there is no photon beam passing via mirror F. According to Danan \emph{et al.} \cite{danan2013asking} there is complete destructive interference of light that is directed towards mirror F. The power spectrum records the frequencies of only three mirrors A, B and C.}  
\end{figure}

\subsection{Beam Splitter 3}

Here we first look at the unitary transformation given by the action of the beam splitter $\hat{B}$, given in equation (1),
\begin{equation}
U_{BS}\left|\gamma\right\rangle_{T}\left|0\right\rangle_{L} = \left|\gamma B_{11}\right\rangle_{T^{\prime}}\left|\gamma B_{21}\right\rangle_{L^{\prime}}
\end{equation}
In the same way as in the case of BS3 in setup 1, we can find $\left|\gamma\right\rangle$, given the states $\left|\gamma B_{11}\right\rangle$ and $\left|\gamma B_{21}\right\rangle$. Proceeding in this way, we find the output states here. The input states are

\begin{equation}
\begin{aligned}
\alpha^{\prime}_{A} = \alpha \sqrt{\frac{1}{3}} e^{-i[(\Psi_{1}-\Phi_{1})/2]} e^{i[(\Psi_{E}(t)-\Phi_{E}(t))/2]} e^{-i[(\Psi_{2}-\Phi_{2})/2]} \\ \times  e^{i[(\Psi_{A}(t)-\Phi_{A}(t))/2]}, \\
\alpha^{\prime}_{B} = \alpha \sqrt{\frac{1}{3}} e^{-i[(\Psi_{1}-\Phi_{1})/2]} e^{i[(\Psi_{E}(t)-\Phi_{E}(t))/2]} e^{i[(\Psi_{2}+\Phi_{2})/2]} \\ \times e^{-i[(\Psi_{B}(t)-\Phi_{B}(t))/2]} e^{i\lambda}  
\end{aligned}
\end{equation}
\begin{equation}
\begin{aligned}
\left| \psi_{in}\right\rangle_{BS3} = \left|\alpha^{\prime}_{A}\right\rangle_{T6} \otimes \left|\alpha^{\prime}_{B}\right\rangle_{L6}
\end{aligned}
\end{equation}
In state L6, we consider an additional phase shift by $e^{i\lambda}$. This is taken into account because of the slight shifting of mirror B for this setup \cite{danan2013asking}. The output state is
\begin{equation}
\begin{aligned}
\left| \psi_{out}\right\rangle_{BS3} = \left| \chi \right\rangle_{d} \otimes \left| 0 \right\rangle_{L7}
\end{aligned}
\end{equation}
with
\begin{equation}
\begin{aligned}
\chi = \alpha\sqrt{\frac{2}{3}}e^{-i[(\Psi_{1}-\Phi_{1})/2]}e^{i[(\Phi_{2}-\Phi_{3}+ \pi)/2]} \times \\ e^{i[(\Psi_{E}(t)-\Phi_{E}(t))/2]}e^{i[(\Psi_{A}(t)-\Phi_{A}(t))/4]} \times \\ e^{-i[(\Psi_{B}(t)-\Phi_{B}(t))/4]} e^{i\lambda}
\end{aligned}
\end{equation}
Here $\Phi_{3}$ is the constant Euler angle associated with the beam splitter BS3. The systems having this state cannot reach the detector D as it emerges from the bottom port of BS3. This is in accordance to ref. \cite{danan2013asking} as shown in FIG. 4. However we take note of one important relation between the phases, which can be used later for finding the final state entering D. This relation is
\begin{equation}
\begin{aligned}
\Psi_{2} + \Psi_{3} = \left[ \frac{\Psi_{A}(t) + \Psi_{B}(t) - \Phi_{A}(t) - \Phi_{B}(t)}{2} - \pi + \lambda \right] 
\end{aligned}
\end{equation}
Equation (46) is obtained while finding the expression for $\chi$. This has been discussed in the \textbf{Appendix}.  

\subsection{Beam Splitter 4}

The system L9 that enters the beam splitter BS4 in FIG. 4 has a non-trivial state. This makes the input state to BS4 as:
\begin{equation}
\begin{aligned}
\left| \psi_{in}\right\rangle_{BS4} = \left| 0 \right\rangle_{T9} \otimes \\ \left|-\alpha \sqrt{\frac{1}{3}}e^{i(\frac{\Psi_{1}+\Phi_{1}}{2})} e^{-i[(\Psi_{C}(t)-\Phi_{C}(t))/2]}\right\rangle_{L9}. 
\end{aligned}
\end{equation}
To find the output state we use relation (46) in our usual beam splitter transformation. Thus the final output state in this setup, from BS4 is
\begin{equation}
\begin{aligned}
\left| \psi_{out}\right\rangle_{BS4} = \left|\alpha_{1}\right\rangle_{g} \otimes \left|\alpha_{final}\right\rangle_{h} 
\end{aligned}
\end{equation}
where,
\[
\alpha_{1}= -\alpha \frac{\sqrt{2}}{3}e^{i(\frac{\Psi_{1}+\Phi_{1}}{2})} e^{i(\frac{\Psi_{4}-\Phi_{4}}{2})}  e^{-i[(\Psi_{C}(t)-\Phi_{C}(t))/2]},
\]
\[
\begin{aligned}
\alpha_{final}= \frac{\alpha}{3} e^{-i(\Psi_{2}+\Psi_{3})} e^{i(\frac{\Psi_{1}+\Phi_{1}}{2})} e^{-i(\frac{\Psi_{4}+\Phi_{4}}{2})} e^{i\lambda} \\ \times e^{i[(\Psi_{A}(t)-\Phi_{A}(t))/2]} e^{i[(\Psi_{B}(t)-\Phi_{B}(t))/2]} \\ \times e^{-i[(\Psi_{C}(t)-\Phi_{C}(t))/2]}.
\end{aligned}
\]
The coherent state $\left|\alpha_{final}\right\rangle_{h}$ of the system \emph{h} reaches the detector D. It corresponds to $\alpha_{final}$ which contains the vibrational information of only three mirrors A, B and C. Also, the phases $\Phi_{1}, \Phi_{2}, \Phi_{3}, \Phi_{4}, \Psi_{1}, \Psi_{2}, \Psi_{3}, \Psi_{4}$ are all constants. The time dependent phases contain the frequencies of their respective mirrors as below:
\begin{equation}
\begin{aligned}
\Phi_{A}(t) = \phi_{0}\sin (2\pi f_{A} t), \\
\Phi_{B}(t) = \phi_{0}\sin (2\pi f_{B} t), \\
\Phi_{C}(t) = \phi_{0}\sin (2\pi f_{C} t), \\
\Phi_{E}(t) = \phi_{0}\sin (2\pi f_{E} t), \\
\Phi_{F}(t) = \phi_{0}\sin (2\pi f_{F} t). 
\end{aligned}
\end{equation}
The same relations hold for phases $\Psi_{i}(t)$ (where \emph{i}=A,B,C,E,F).
The final state which contains some or all of these time dependant phases will be measured by the detector D, by the corresponding frequencies of those phases.

Relation (40) gives the final output for \textbf{setup 1}, and (48) gives the final output for \textbf{setup 2}.

\section{POWER SPECTRUM ANALYSIS OF THE FINAL OUTPUT STATES}

Here, we begin with a quantum Power Spectral Density (PSD). This is a spectral function which gives the intensity of a time-dependent quantum mechanical operator $\hat{a}(t)$ for a given frequency $\omega$. It is defined as \cite{clerk2010introduction}:
\begin{equation}
\begin{aligned}
S_{aa}(\omega) = \int\limits_{-\infty}^{\infty}R_{aa}(t)e^{i\omega t}dt \\
= \int\limits_{-\infty}^{\infty}\left\langle \hat{a}(t)\hat{a}(0)\right\rangle e^{i\omega t}dt
\end{aligned}
\end{equation}
Here $R_{aa}(t) = \left\langle \hat{a}(t)\hat{a}(0)\right\rangle$ is the auto-correlation function for $\hat{a}(t)$. The coherent state entering into the detector D will be analyzed using the above function.
     
Accordingly, the power spectrum analysis of the final states obtained from setups 1 and 2 have been done using the following power spectral function, given in ref. \cite{hauer2015nonlinear}:
\[
\begin{aligned}
S_{xx}(\omega) = x^{2}_{zpf} \int\limits_{-\infty}^{\infty}[(\alpha)^{2}e^{-i\omega_{0}t} + (\alpha^{*})^{2}e^{i\omega_{0}t} + (\alpha^{*}\alpha)e^{i\omega_{0}t} + \\ (1+\alpha^{*}\alpha)e^{-i\omega_{0}t}]e^{i\omega t} dt \\
= 2\pi x^{2}_{zpf} [(\alpha)^{2}\delta(\omega - \omega_{0}) + (\alpha^{*})^{2}\delta(\omega + \omega_{0}) + \\ |\alpha|^{2}\delta(\omega + \omega_{0}) + (1 + |\alpha|^{2})\delta(\omega - \omega_{0})]
\end{aligned}
\]
We know the $\alpha$ of the final coherent states $\left|\alpha\right\rangle_{h}$ for setups 1 and 2. Putting those values in, we obtain the final power spectra for the two setups separately. Here $\omega_{0}$ and $-\omega_{0}$ are the electromagnetic radiation frequencies associated with a single quantum state. $x_{zpf} = \sqrt{\frac{\hbar}{2m\omega_{0}}}$ is the zero-point fluctuation constant.

\subsection{Power Spectrum of Output State from Setup 1}

Putting $\alpha$ for setup 1, into the power spectral function above we have,
\[
\begin{aligned}
S^{(1)}_{xx}(\omega) = 2\pi x^{2}_{zpf} [((\alpha)^{2} e^{i2\kappa} e^{i[(\Psi_{E}(t)-\Phi_{E}(t))/2]} \\ \times e^{i[(\Psi_{F}(t)-\Phi_{F}(t))/2]} e^{i[(\Psi_{A}(t)-\Phi_{A}(t))/4]} \\ \times e^{-i[(\Psi_{B}(t)-\Phi_{B}(t))/4]} e^{-i[(\Psi_{C}(t)-\Phi_{C}(t))/2]} \\ \times \delta(\omega - \omega_{0})) \\  + ((\alpha^{*})^{2} e^{-i2\kappa} e^{-i[(\Psi_{E}(t)-\Phi_{E}(t))/2]} \\ \times e^{-i[(\Psi_{F}(t)-\Phi_{F}(t))/2]} e^{-i[(\Psi_{A}(t)-\Phi_{A}(t))/4]} \\ \times e^{i[(\Psi_{B}(t)-\Phi_{B}(t))/4]} e^{i[(\Psi_{C}(t)-\Phi_{C}(t))/2]} \\ \times \delta(\omega + \omega_{0})) \\ + |\alpha|^{2}\delta(\omega + \omega_{0}) + (1 + |\alpha|^{2})\delta(\omega - \omega_{0})].
\end{aligned}
\]
We observe that in this spectrum, we must get the frequencies corresponding to mirrors A,B,C,E and F as in equation (49). This is easily seen from the expression for $S^{(1)}_{xx}(\omega)$ where the time dependent phases of all the five mirrors are present. This is in agreement with the experimental result of Danan \emph{et al.} \cite{danan2013asking}. Here $\kappa$ is a constant phase. 

\subsection{Power Spectrum of Output State from Setup 2}

Now we put $\alpha$ for setup 2 into the power spectral function:
\[
\begin{aligned}
S^{(2)}_{xx}(\omega) = 2\pi x^{2}_{zpf} [ (\frac{\alpha^{2}}{9} e^{i2\kappa^{\prime}} e^{i[\Psi_{A}(t)-\Phi_{A}(t)]} \\ \times e^{i[\Psi_{B}(t)-\Phi_{B}(t)]} e^{-i[\Psi_{C}(t)-\Phi_{C}(t)]} \\ \times \delta(\omega - \omega_{0})) \\ + (\frac{\alpha^{*2}}{9} e^{-i2\kappa^{\prime}} e^{-i[\Psi_{A}(t)-\Phi_{A}(t)]} \\ \times e^{-i[\Psi_{B}(t)-\Phi_{B}(t)]} e^{i[\Psi_{C}(t)-\Phi_{C}(t)]} \\ \times \delta(\omega - \omega_{0}))  \\ + |\alpha|^{2}\delta(\omega + \omega_{0}) + (1 + |\alpha|^{2})\delta(\omega - \omega_{0}) ]. 
\end{aligned}
\]
From this expression for $S^{(2)}_{xx}(\omega)$, we see that the power spectrum depends on the frequencies of mirrors A, B and C only (from equation (49)). Here $\kappa^{\prime}$ is a constant phase. This is exactly what Danan \emph{et al.} observed in their experiment \cite{danan2013asking}. 

Our quantum mechanical state vector calculations with coherent states have proved that this observation is indeed expected. Thus our simple coherent state approach is enough to explain the outcomes of this experiment. Although we have used a quantum mechanical approach, it is important to note that we have used a coherent state and calculated its evolution for both the setups (1 and 2). This points out the classical nature of the photon beams used in this experiment. As any single-mode classical state is a convex mixture of (single mode) coherent states -- via the Glauber-Sudarshan P-distribution \cite{sudarshan1963equivalence, glauber1963coherent} -- our analysis shows that the conclusion of the experiment in ref. \cite{danan2013asking} follows also from any single-mode classical state in the input mode T1 (of FIG. 3 as well as FIG. 4). Hence the observations of this experiment can also be expected to be explained using classical light -- as has been done in ref. \cite{saldanha2014interpreting}.

\section{CONCLUSION}

We have provided here (without using the two-state vector formalism of ref. \cite{aharonov1964time}), a fully quantum mechanical description of the experiment in ref. \cite{danan2013asking} using any classical (in the sense of Quantum Optics) input state of a single-mode radiation field, and qualitatively established similar types of functional dependence of the power spectra on the oscillation frequencies of the mirrors used in the nested interferometric experiment of ref. \cite{danan2013asking}. We believe that the power spectra we obtained here do match quantitatively also with those of ref. \cite{danan2013asking}. We do hope that our analysis here does help in understanding the foundational issues related to paths of microscopic systems --- as discussed in ref. \cite{danan2013asking}.

\section{Acknowledgements}

The work of AR has been supported by the Summer Internship Programme in Physics of the Institute of Mathematical Sciences (IMSc.), Chennai, India and the Kishore Vaigyanik Protsahan Yojana (KVPY) Fellowship Programme of the Department of Science \& Technology, Government of India. Most part of the work was done when AR was a final year BS-MS 5-year Dual Degree course student (in Physics) at IISER-Bhopal, from where he graduated recently. SG would like to thank Sandeep K. Goyal for useful discussion on ref. \cite{danan2013asking}.  

\appendix*
\section{Proof of Eqn. (46)}
We use equations (41), (42), and (43), to find initially, two separate expressions for $\gamma$ as shown below:
\begin{equation}
\begin{aligned}
\gamma \sqrt{\frac{1}{2}} e^{i[(\Psi_{3}+\Phi_{3})/2]} = \alpha \sqrt{\frac{1}{3}} e^{-i[(\Psi_{1}-\Phi_{1})/2]} e^{i[(\Psi_{E}(t)-\Phi_{E}(t))/2]} \\ \times e^{-i[(\Psi_{2}-\Phi_{2})/2]} e^{i[(\Psi_{A}(t)-\Phi_{A}(t))/2]}
\end{aligned}
\end{equation}
\begin{equation}
\begin{aligned}
-\gamma \sqrt{\frac{1}{2}} e^{-i[(\Psi_{3}-\Phi_{3})/2]} = \alpha \sqrt{\frac{1}{3}} e^{-i[(\Psi_{1}-\Phi_{1})/2]} \times \\ e^{i[(\Psi_{E}(t)-\Phi_{E}(t))/2]} e^{i[(\Psi_{2}+\Phi_{2})/2]} \times \\ e^{-i[(\Psi_{B}(t)-\Phi_{B}(t))/2]} e^{i\lambda}
\end{aligned}
\end{equation}
Next, we obtain the two different expressions for $\gamma$ from equations (A.1) and (A.2). This process is trivial, since we know the above equations. 

Now, we need to equate these two separate expressions for $\gamma$. Doing this, we obtain the following relation:
\begin{equation}
\begin{aligned}
\alpha \sqrt{\frac{2}{3}} e^{-i[(\Psi_{1}-\Phi_{1})/2]} e^{i[(\Psi_{E}(t)-\Phi_{E}(t))/2]} e^{-i[(\Psi_{2}-\Phi_{2})/2]} \times \\ e^{i[(\Psi_{A}(t)-\Phi_{A}(t))/2]} e^{-i[(\Psi_{3}+\Phi_{3})/2]} = \\
\alpha \sqrt{\frac{2}{3}} e^{-i[(\Psi_{1}-\Phi_{1})/2]} e^{i[(\Psi_{E}(t)-\Phi_{E}(t))/2]} e^{i[(\Psi_{2}+\Phi_{2})/2]} \times \\ e^{-i[(\Psi_{B}(t)-\Phi_{B}(t))/2]} e^{i[(\Psi_{3}-\Phi_{3})/2]} e^{i\lambda} e^{i\pi}
\end{aligned}
\end{equation}
On simplifying equation (A.3), and solving for $\Psi_{2} + \Psi_{3}$, we obtain equation (46):
\[
\begin{aligned}
\Psi_{2} + \Psi_{3} = \left[ \frac{\Psi_{A}(t) + \Psi_{B}(t) - \Phi_{A}(t) - \Phi_{B}(t)}{2} - \pi + \lambda \right] 
\end{aligned}
\]

\end{document}